\documentclass[12pt]{article}

\usepackage[margin=1in]{geometry}
\usepackage{amsmath,amssymb,amsthm}
\usepackage{booktabs}
\usepackage{graphicx}
\usepackage{enumitem}
\usepackage{setspace}
\usepackage{xcolor}
\usepackage{natbib}
\usepackage{hyperref}
\usepackage{tikz}
\usepackage{pgfplots}
\usepackage{comment}
\usepackage{authblk}

\pgfplotsset{compat=1.18}
\usetikzlibrary{arrows.meta,positioning,calc}

\hypersetup{colorlinks=true,linkcolor=blue,citecolor=blue,urlcolor=blue}
\theoremstyle{plain}
\newtheorem{proposition}{Proposition}
\newtheorem{corollary}{Corollary}
\theoremstyle{remark}

\title{Pseudo-IPD Reconstruction from Published Kaplan--Meier Curves under Unidentified Within-Interval Censoring}

\author[1]{Yi-Cheng Tai, PhD}
\author[2]{Weijing Wang, PhD\textsuperscript{*}}
\author[3]{Martin T. Wells, PhD\textsuperscript{*}}

\affil[1]{Department of Statistics, National Chengchi University,
Taipei, Taiwan}
\affil[2]{Institute of Statistics, National Yang Ming Chiao Tung University,
Hsinchu City, Taiwan}
\affil[3]{Department of Statistics and Data Science, Cornell University,
New York, USA}

\affil[*]{Corresponding authors}

\date{}

\begin{document}
\makeatletter
\let\KMsaved@maketitle\maketitle
\let\KMsaved@@maketitle\@maketitle
\let\KMsaved@author\@author
\makeatother
\maketitle
\doublespacing

\begin{abstract}
Reconstruction of pseudo-individual patient data (pseudo-IPD) from published
Kaplan--Meier curves and number-at-risk tables is widely used to support
secondary survival analyses when individual-level data are unavailable.
However, the reconstruction is generally not unique because the published
information does not determine how censoring is distributed within each
risk-table interval.
We study the non-identifiability that remains after interval-level event and
censoring totals have been reconstructed. Different within-interval censoring
placements can yield similarly good Kaplan--Meier agreement but different
risk-set and censoring patterns, with possible effects on subsequent analyses.
We characterize this non-identifiability, introduce a systematic allocation
scheme, and assess its impact on reconstructed pseudo-IPD and downstream
analyses.
Simulation studies were conducted to examine how within-interval censoring
allocation affects pseudo-IPD reconstruction under different risk-table
resolutions and censoring settings. The proposed approach was also illustrated using published data from
CheckMate~067.
Our analysis suggests that considering multiple reconstructions provides a
practical way to assess the impact of this non-identifiability on downstream
analyses.

\end{abstract}


\clearpage

\section{Introduction and Problem Formulation}
\label{sec:introduction-setting}

Published reports of randomized clinical trials with time-to-event outcomes
often provide Kaplan--Meier (KM) curves and numbers at risk at selected time
points, but not the original individual-level observations. Reconstruction of pseudo-individual patient data (pseudo-IPD) from these summaries has therefore become useful for secondary survival analyses and methodological comparisons. Early approaches estimated interval-level event and censoring counts from digitized Kaplan--Meier curves and reported numbers at risk \citep{Parmar1998, Tierney2007, Hoyle2011}. \citet{Guyot2012EnhancedSecondaryAnalysis} developed
the widely used iKM reconstruction algorithm, and subsequent work has evaluated
or extended this approach \citep{Wan2015Review,Liu2021IPDfromKM}. Related work
has also developed procedures and software to extract survival information
from published Kaplan--Meier curves \citep{Zhang2024SurvdigitizeR, Zhao2025, Ke2026StochIKM}. When
the censoring times are marked on the published curve,
\citet{Rogula2022MarkedCensoring} showed how those marks can be incorporated
directly. More recently, \citet{Titman2026QP} formulated pseudo-IPD
reconstruction as a quadratic-programming problem, with event and censoring
counts chosen subject to constraints from reported numbers at risk, total event
counts, and marked censoring times. The approach was also extended to
competing-risks data.

The present paper takes a different perspective on the inverse problem underlying pseudo-IPD reconstruction. Obtaining a pseudo-IPD reconstruction does not imply that the underlying individual-level data are uniquely determined. A published Kaplan--Meier curve together with numbers at risk reported on a discrete time grid generally does not determine a unique set of individual-level survival data. Specifically,
 reconstruction involves two stages. The first recovers interval-level
event and censoring totals from the Kaplan--Meier curve and risk table. The
second determines how those totals are distributed within each reporting
interval. Existing reconstruction methods, including
\citet{Guyot2012EnhancedSecondaryAnalysis}, make choices at both stages to
construct pseudo-IPD. The within-interval censoring allocation is typically
fixed as part of that construction and is not examined separately.

Our objective is to assess how the remaining non-identifiability in within-interval censoring allocation affects pseudo-IPD reconstruction and subsequent survival analyses after the interval-level totals have been fixed. We examine the sensitivity of reconstructed pseudo-IPD and subsequent analyses to alternative censoring allocations.
Our analysis proceeds in two parts. First, we examine how within-interval censoring allocation affects the reconstructed pseudo-IPD. Second, we introduce working censoring profiles to represent alternative within-interval allocations and evaluate their effects on the reconstruction and subsequent analyses. We also investigate the role of risk-table resolution. The framework is assessed through simulation studies and an application to published Kaplan--Meier data from CheckMate~067. The simulations evaluate reconstruction behavior under controlled settings, while data analysis illustrates the practical impact of alternative censoring allocations in a published trial.

\subsection{Notation and published KM information}

Let \(T\) denote the event time and \(C\) the censoring time, and define
\[
X=\min(T,C),
\qquad
\delta=I(T\le C),
\]
where \(X\) is the observed time and \(\delta\) is the event indicator. Let
\(S(t)=P(T>t)\) denote the survival function. Under independent censoring, if
 individual observations \((X_i,\delta_i)\), \(i=1,\ldots,n\), were
available, \(S(t)\) could be estimated by the Kaplan--Meier estimator.
\[
\widehat S_{\mathrm{KM}}(t)
=
\prod_{u\le t}
\left\{
1-
\frac{\sum_{i=1}^n I(X_i=u,\delta_i=1)}
{\sum_{i=1}^n I(X_i\ge u)}
\right\}.
\]
For reconstruction from a published Kaplan--Meier curve, digitization provides numerical time and survival coordinates \citep{Liu2021IPDfromKM,Zhang2024SurvdigitizeR}. We denote the resulting
digitized curve by \(\widehat S_{\mathrm{dig}}(t)\), which approximates the
published curve \(\widehat S_{\mathrm{KM}}(t)\). These coordinates, together
with the reported numbers at risk, provide the basic information used for
pseudo-IPD reconstruction.
Suppose that the numbers at risk are reported at
\[
t_0<t_1<\cdots<t_K,
\]
and define the corresponding risk-table intervals by
\[
I_k=[t_k,t_{k+1}),
\qquad
k=0,\ldots,K-1.
\]
Let
\[
Y_k=\sum_{i=1}^n I(X_i\ge t_k),
\qquad
k=0,\ldots,K,
\]
denote the reported number at risk at \(t_k\).
The digitized Kaplan--Meier curve is usually available at a much finer time
resolution than the number-at-risk table, so several survival drops may occur
within a single interval \(I_k\). Let
\begin{equation}
u_{k1}<u_{k2}<\cdots<u_{kJ_k},
\qquad
u_{kj}\in I_k,
\label{eq:candidate-drop-times}
\end{equation}
denote the candidate drop times in \(I_k\), where \(J_k\) is the number of
digitized drops. For each \(u_{kj}\), let
\[
\widehat S_{kj}^{-}
=
\widehat S_{\mathrm{dig}}(u_{kj}^{-}),
\qquad
\widehat S_{kj}
=
\widehat S_{\mathrm{dig}}(u_{kj}),
\]
denote the digitized survival levels immediately before and after the drop,
respectively, and define the corresponding conditional drop fraction by
\begin{equation}
\widehat h_{kj}
=
1-\frac{\widehat S_{kj}}{\widehat S_{kj}^{-}},
\qquad
j=1,\ldots,J_k.
\label{eq:drop-fraction}
\end{equation}
Thus, \(u_{kj}\) and \(\widehat h_{kj}\) describe the timing and relative
magnitude of the digitized survival drops within \(I_k\).

Let \(D_k\) and \(C_k\) denote the reconstructed numbers of events and censored
observations, respectively, in \(I_k\). They satisfy
\begin{equation}
Y_k-Y_{k+1}=D_k+C_k.
\label{eq:risk-balance}
\end{equation}
The digitized curve provides information about the timing and magnitude of
survival drops within the interval, whereas the risk table constrains the total
reduction in the risk set between its endpoints. Even when \(D_k\) and \(C_k\)
are fixed, however, the published information does not determine how the
censoring total \(C_k\) is distributed within \(I_k\). Related nonuniqueness has
also been recognized in optimization-based reconstruction
\citep{Titman2026QP}. This unresolved allocation may affect the reconstructed
risk sets and event allocation and, consequently, subsequent analyses. We
therefore examine several working allocations to assess variation in the
reconstructed pseudo-IPD and downstream results.

\subsection{Guyot-style reconstruction and censoring allocation}

We begin with the reconstruction of
\citet{Guyot2012EnhancedSecondaryAnalysis}, whose sequential formulation makes
the role of within-interval censoring placement explicit. A different
formulation was proposed by \citet{Titman2026QP}, who expressed event and
censoring allocation as a quadratic-programming problem under constraints from
the published information.
At a candidate drop time \(u_{kj}\), the Kaplan--Meier relation gives
\[
d_{kj}=Y(u_{kj})\,\widehat h_{kj},
\]
where \(Y(u_{kj})\) is the number at risk immediately before the drop. The
digitized curve determines \(\widehat h_{kj}\), but \(Y(u_{kj})\) depends on
events and censoring that have occurred earlier in the interval. Different
within-interval censoring placements can therefore lead to different event
counts at the same candidate drop time.

In the Guyot reconstruction, the interval censoring total is estimated from
the digitized Kaplan--Meier curve and the reported numbers at risk. Working
censoring times are then placed approximately evenly within the interval, and
the candidate drop times are processed sequentially to reconstruct the event
counts. The censoring total is adjusted, if necessary, until the reported
number at risk at \(t_{k+1}\) is reproduced. The resulting interval totals
\(D_k\) and \(C_k\) satisfy
\[
Y_k-Y_{k+1}=D_k+C_k.
\]
Detailed calculations for the Guyot-style first-stage reconstruction are given in Supplementary Section~S2.

The approximately even placement of censoring times in the Guyot reconstruction
is a working assumption. The published Kaplan--Meier curve and risk table do
not uniquely determine the observed times of censored subjects within a
risk-table interval. This remaining within-interval allocation is the focus of
the method developed below.

\section{Within-interval censoring allocation}
\label{sec:refinement}

We take the quantities obtained from the first-stage reconstruction as fixed.
For each risk-table interval \(I_k\), these include the interval-level totals
\((D_k,C_k)\), the candidate Kaplan--Meier drop times, and the digitized drop
fractions. The remaining unspecified component is the distribution of the
\(C_k\) censoring observations within the interval.

\subsection{Working censoring profiles and event-grid allocation}

Set \(u_{k0}=t_k\) and \(u_{k,J_k+1}=t_{k+1}\). Together with the candidate
drop times, these points partition \(I_k\) into
\begin{equation}
L_{kj}=[u_{k,j-1},u_{kj}),
\qquad
j=1,\ldots,J_k+1.
\label{eq:local-censoring-segments}
\end{equation}
We represent alternative within-interval censoring placements by a
nondecreasing working profile \(q_{k,m}(t)\), with
\[
q_{k,m}(t_k)=0,
\qquad
q_{k,m}(t_{k+1})=1.
\]
The primary family is
\begin{equation}
q_{k,m}(t)
=
\left(
\frac{t-t_k}{t_{k+1}-t_k}
\right)^{\alpha_m},
\qquad
\alpha_m>0,
\label{eq:working-profile}
\end{equation}
with \((\alpha_1,\alpha_2,\alpha_3)=(0.5,1,2)\), representing early, linear,
and late censoring placement. For convenience and comparability across intervals, the same value of \(\alpha_m\) is used across risk-table intervals for each profile. These profiles are working allocations and are not models for the true censoring distribution.

For segment \(L_{kj}\), define the real-valued censoring allocation
\begin{equation}
\Delta_{kj,m}^{*}
=
C_k
\left\{
q_{k,m}(u_{kj})-q_{k,m}(u_{k,j-1})
\right\},
\qquad
j=1,\ldots,J_k+1.
\label{eq:censor-target-increments}
\end{equation}
The resulting segment allocations sum to \(C_k\). They are converted to
integer counts \(c_{kj,m}\) using a deterministic largest-remainder rule, with
\[
\sum_{j=1}^{J_k+1}c_{kj,m}=C_k.
\]
The deterministic rule of the largest remaining rule and its finite-sample bounds are given in
the Supplementary Section~S3.
For each candidate drop time \(u_{kj}\), define the cumulative censoring count
before that drop by
\begin{equation}
M_{kj,m}
=
\sum_{\ell=1}^{j}c_{k\ell,m},
\qquad
j=1,\ldots,J_k.
\label{eq:cumulative-censor-count}
\end{equation}
Censoring in the final segment
\(L_{k,J_k+1}=[u_{kJ_k},t_{k+1})\) occurs after the last candidate drop and
does not contribute to \(M_{kj,m}\). Define
\begin{equation}
\widetilde Y_{kj,m}
=
Y_k-M_{kj,m}.
\label{eq:intermediate-risk}
\end{equation}
Unlike the sequential Guyot reconstruction, $\widetilde Y_{kj,m}$ nets out censoring only, not earlier events in $I_k$; it serves as a weight for distributing the fixed total $D_k$ and is not intended to reproduce the number at risk immediately before each drop.

\subsection{Event allocation}

Using the digitized drop fractions in Equation~\eqref{eq:drop-fraction}, define
\begin{equation}
w_{kj,m}
=
\widetilde Y_{kj,m}\widehat h_{kj},
\qquad
j=1,\ldots,J_k,
\label{eq:event-weight}
\end{equation}
and let
\[
W_{k,m}
=
\sum_{j=1}^{J_k}w_{kj,m}.
\]
When \(W_{k,m}>0\), allocate
the fixed interval-level event total \(D_k\) according to
\begin{equation}
d_{kj,m}^{*}
=
D_k\frac{w_{kj,m}}{W_{k,m}},
\qquad
j=1,\ldots,J_k.
\label{eq:event-allocation}
\end{equation}
The real-valued allocations are converted to integer event counts \(d_{kj,m}\)
using the same deterministic largest-remainder rule described in Supplementary
Section~S3, with
\[
\sum_{j=1}^{J_k}d_{kj,m}=D_k.
\]
Hence every working profile satisfies the same interval endpoint constraint,
\[
Y_k
-
\sum_{j=1}^{J_k}d_{kj,m}
-
\sum_{j=1}^{J_k+1}c_{kj,m}
=
Y_{k+1}.
\]

For profile \(m\), the reconstructed pseudo-IPD contain \(d_{kj,m}\) events at
the candidate drop times \(u_{kj}\) and \(c_{kj,m}\) censored observations
assigned within the segments \(L_{kj}\). The event allocation depends on the
censoring configuration through the cumulative counts \(M_{kj,m}\) at the
candidate drop times. Once the segment-specific counts are fixed, the exact
censoring times within a segment do not affect these cumulative counts. This
property is formalized in Section~\ref{sec:theoretical-properties}; proofs of the
theoretical results are given in Supplementary Section~S4.

\section{Sensitivity analysis across working censoring profiles}
\label{sec:censoring-profile-sensitivity}

Each working profile produces a reconstructed pseudo-IPD dataset
\begin{equation}
\mathcal D_m^{\mathrm{pseudo}}
=
\left\{
\left(X_{im}^{\mathrm{pseudo}},\delta_{im}^{\mathrm{pseudo}}\right):
i=1,\ldots,n
\right\}.
\label{eq:pseudo-ipd-profile}
\end{equation}
The sensitivity analysis examines how alternative within-interval censoring
allocations affect the reconstructed data and analyses based on those data.
We consider four quantities: the cumulative censoring process, the
number-at-risk process, the refitted Kaplan--Meier curve, and downstream
survival summaries.

The cumulative censoring process is
\begin{equation}
N_m^C(t)
=
\sum_i
I\left(
X_{im}^{\mathrm{pseudo}}\le t,
\delta_{im}^{\mathrm{pseudo}}=0
\right).
\label{eq:censoring-process-profile}
\end{equation}
Differences in \(N_m^C(t)\) describe how the timing of censoring changes across
working profiles while the interval-level censoring totals remain fixed. This
quantity is also relevant to procedures that depend explicitly on the
censoring distribution, such as inverse-probability-of-censoring weighting.
Changes in censoring placement also alter the subjects remaining under
observation. The corresponding number-at-risk process is
\begin{equation}
Y_m(t)
=
\sum_i
I\left(
X_{im}^{\mathrm{pseudo}}\ge t
\right).
\label{eq:risk-process-profile}
\end{equation}
Variation in \(Y_m(t)\) describes differences in the reconstructed risk sets,
which enter the event reconstruction and subsequent survival analyses,
including Kaplan--Meier estimation, log-rank comparisons, and Cox regression.

We also examine whether differences in censoring and risk-set trajectories are
reflected in the reconstructed survival curve. Let \(\widehat S_m(t)\) denote
the Kaplan--Meier curve refitted from
\(\mathcal D_m^{\mathrm{pseudo}}\). Similar refitted curves across working
profiles can occur even when the corresponding censoring and risk-set
trajectories differ.
Finally, we examine quantities used in downstream analyses. One example is the
restricted mean survival time up to \(\tau\),
\begin{equation}
\widehat{\mathrm{RMST}}_m(\tau)
=
\int_0^\tau \widehat S_m(t)\,dt.
\label{eq:rmst-profile}
\end{equation}
In multi-arm settings, survival contrasts, RMST contrasts, and hazard ratios
may likewise be compared across working profiles. Such comparisons show
whether differences arising from within-interval censoring allocation affect
the conclusions of subsequent analyses.

Sensitivity is assessed separately for each quantity across working profiles
that share the same interval-level totals \((D_k,C_k)\), candidate drop times,
and digitized drop fractions. Because the quantities differ in interpretation
and scale, variation is compared across profiles within a given measure, not
across different measures. Greater across-profile variation indicates greater
sensitivity to the unidentified within-interval censoring allocation.
The simulation evaluates this sensitivity for the refitted Kaplan--Meier
curve, RMST, the number-at-risk process, and the cumulative censoring process.
Details of the reconstruction discrepancies and the across-profile sensitivity
range are given in Supplementary Section~S5.2. The drop-density and
allocation-relevance diagnostics used to examine the role of risk-table
resolution are defined in Supplementary Section~S5.3.

\section{Theoretical Properties}
\label{sec:theoretical-properties}

The theoretical analysis addresses three questions. First, we determine when
different within-interval censoring allocations lead to the same event
reconstruction. Second, we give an explicit count of the possible
within-interval censoring patterns that satisfy a fixed interval-level
censoring total. Third, we distinguish variation caused by within-interval
censoring placement from numerical error introduced by integer allocation.

For fixed interval-level totals, candidate drop times, and digitized drop
fractions, the event allocation depends on the cumulative amount of censoring
before each candidate drop time. For interval \(I_k\), define
\[
\mathbf M_{k,m}
=
(M_{k1,m},\ldots,M_{kJ_k,m}),
\]
where \(M_{kj,m}\) is the cumulative censoring count defined in
Equation~\eqref{eq:cumulative-censor-count}. This vector contains the
within-interval censoring information that enters the event-allocation rule.

\subsection{Event-grid equivalence}

We characterize conditions under which two different within-interval censoring
allocations lead to the same reconstructed event counts at the candidate drop
times.

\begin{proposition}[Event-grid equivalence]
\label{prop:event-grid-equivalence}
Fix a risk-table interval \(I_k\), the interval-level totals
\((D_k,C_k)\), the candidate drop times, and the digitized drop fractions.
If two working allocations \(m\) and \(m'\) have the same cumulative
censoring counts at every candidate drop time,
\[
\mathbf M_{k,m}
=
\mathbf M_{k,m'},
\]
then they produce the same real-valued event allocation,
\[
d_{kj,m}^{*}=d_{kj,m'}^{*},
\qquad
j=1,\ldots,J_k,
\]
and, under the same deterministic integer-allocation rule, the same integer
event counts.
\end{proposition}

Conditional on the fixed first-stage information, event reconstruction depends
on the cumulative censoring counts at the candidate drop times. Different
censoring placements that agree on these counts therefore lead to the same
event allocation.
This also implies that, once the number of censoring observations in each
segment is fixed, their exact locations within the segment do not affect the
event reconstruction.

\begin{corollary}[Within-segment invariance]
\label{cor:within-segment-invariance}
If the segment-specific censoring counts \(c_{kj,m}\) are fixed, changing the
exact censoring times within any segment \(L_{kj}\) does not change the
reconstructed event allocation.
\end{corollary}
Proofs of Proposition~\ref{prop:event-grid-equivalence} and
Corollary~\ref{cor:within-segment-invariance} are given in
Supplementary Section~S4.1.

\subsection{Number of possible censoring patterns}

We next quantify the number of within-interval censoring allocations that are
consistent with a fixed interval-level censoring total \(C_k\).

\begin{proposition}[Number of possible censoring patterns]
\label{prop:number-compatible-paths}
If no additional within-interval censoring information is available, the number
of segment-specific allocations satisfying
\[
\sum_{j=1}^{J_k+1} c_{kj}=C_k
\]
is
\[
\binom{C_k+J_k}{J_k}.
\]
\end{proposition}

This number increases with both \(C_k\) and \(J_k\). Hence, intervals with more
censoring observations or more candidate drop times allow a larger number of
within-interval censoring patterns that satisfy the same total. These patterns
can differ in their cumulative censoring counts at the candidate drop times
and can therefore lead to different event allocations.
A proof of Proposition~\ref{prop:number-compatible-paths} is given in
Supplementary Section~S4.2.

\subsection{Preservation of interval-level constraints}

Across different within-interval censoring allocations, we examine what remains
fixed when the interval-level event and censoring totals are preserved.

\begin{proposition}[Endpoint preservation]
\label{prop:endpoint-preservation}
For a fixed risk-table interval \(I_k\), any working allocation satisfying
\[
\sum_{j=1}^{J_k}d_{kj,m}=D_k,
\qquad
\sum_{j=1}^{J_k+1}c_{kj,m}=C_k,
\]
with
\[
D_k+C_k=Y_k-Y_{k+1},
\]
also satisfies
\[
Y_k
-
\sum_{j=1}^{J_k}d_{kj,m}
-
\sum_{j=1}^{J_k+1}c_{kj,m}
=
Y_{k+1}.
\]
\end{proposition}

Proposition~\ref{prop:endpoint-preservation} shows that different
within-interval allocations can reproduce the same reported number at risk at
\(t_{k+1}\). Agreement with the interval endpoint alone therefore does not
identify a unique reconstruction.
A proof of Proposition~\ref{prop:endpoint-preservation} is given in
Supplementary Section~S4.3.

The variation described by these results arises from differences in
within-interval censoring allocation. Numerical discrepancies introduced when
the real-valued censoring and event allocations are converted to integer
counts are considered separately. The deterministic integer-allocation rule
and finite-sample bounds for these discrepancies are given in Supplementary
Section~S3.

\section{Simulation Analysis}
\label{sec:simulation}

\subsection{Simulation design}
\label{subsec:simulation-design}

The simulation was designed to study sensitivity to within-interval censoring
allocation while separating it from error in the reconstruction of the
interval-level totals. We considered the Guyot-style even allocation, a linear
profile, and early and late profiles with \(\alpha=0.5\) and \(2\),
respectively. For each simulated dataset, the fitted Kaplan--Meier curve and
corresponding number-at-risk table were treated as the published information.
Candidate drop times \(u_{kj}\) and drop fractions \(\widehat h_{kj}\) were
obtained directly from the fitted curve, so no additional curve-digitization
error was introduced.

Risk-table grid widths of 3, 6, and 12 months were used. Two settings were
considered for the interval-level totals \((D_k,C_k)\). In the oracle-total
setting, the true interval totals were taken from the simulated observed data.
In the practical-total setting, they were reconstructed from the Kaplan--Meier
curve and number-at-risk table using the method of
\citet{Guyot2012EnhancedSecondaryAnalysis}. Within each dataset and grid width,
all censoring allocations were applied to the same \((D_k,C_k)\) and the same
candidate drop-time grid.

Event times were generated from Weibull distributions with scale parameter 12
and shape parameters \(0.8\), \(1.2\), and \(2.0\), and censoring was generated
under uniform, exponential, Weibull, administrative, and mixture mechanisms.
Each dataset contained 300 subjects, with 500 Monte Carlo replications for each
event-time and censoring setting. Reconstruction performance was evaluated
using the refitted Kaplan--Meier curve, RMST, the number-at-risk process, and
the cumulative censoring process. The main simulation settings are summarized
in Table~\ref{tab:simulation-settings}. Exact data-generating distributions and
the construction of the risk-table grids are given in Supplementary
Section~S5.1.

\subsection{Sensitivity measures}
\label{subsec:simulation-sensitivity}

For each reconstructed pseudo-IPD dataset, four quantities were compared with
their counterparts from the original simulated observed data: the
Kaplan--Meier curve, RMST, the number-at-risk process, and the cumulative
censoring process. RMST was evaluated up to \(\tau\), defined as the 90th
percentile of the observed follow-up times in each simulated dataset, with the
same \(\tau\) used for the original data and all reconstructions of that
dataset. The discrepancy for a given reconstruction summary is denoted by
\(Q_{a,m}^{(g)}\), where \(g\) denotes the risk-table grid width, \(a\) the
oracle- or practical-total setting, and \(m\) the allocation rule. For the
Kaplan--Meier curve, number-at-risk process, and cumulative censoring process,
the discrepancy was based on the difference between the reconstructed and
original observed quantities; for RMST, it was the absolute difference between
the reconstructed and original observed RMST. Details of these calculations,
including the evaluation time points for the process-based discrepancies, are
given in Supplementary Section~S5.2.

Sensitivity was assessed across allocation rules \(m\) while holding the other
reconstruction inputs fixed. For a given reconstruction summary \(Q\), we
summarized this variation by
\[
R_{Q,a}^{(g)}
=
\max_{m\in\mathcal M} Q_{a,m}^{(g)}
-
\min_{m\in\mathcal M} Q_{a,m}^{(g)}.
\]
Larger values of \(R_{Q,a}^{(g)}\) indicate greater sensitivity to the
within-interval censoring allocation. Because the reconstruction summaries are
on different scales, \(R_{Q,a}^{(g)}\) is compared across settings for the same
\(Q\), not across different summaries.

\subsection{Sensitivity results}
\label{subsec:simulation-sensitivity-results}

Sensitivity generally increased as the risk-table grid became coarser.
Table~\ref{tab:simulation-grid-ranges} reports the mean sensitivity range for
each reconstruction summary. For each simulated dataset, the range was
calculated across the four allocation rules and then averaged over the
simulation settings and Monte Carlo replications. Because the four summaries
are on different scales, they are interpreted separately. Under oracle totals,
sensitivity increased consistently as the risk-table grid became coarser for
all four reconstruction summaries. Under practical totals, the same pattern
was clear for the risk-set and censoring processes and was also present for
RMST, while the survival-curve MAE changed little across grid widths. The
oracle- and practical-total results were similar for the finest grid but
differed more as the grid became coarser. The smaller sensitivity observed
under practical totals at coarse grids should not be interpreted as greater
reconstruction accuracy, since reconstruction of \((D_k,C_k)\) can reduce the
number of intervals in which alternative censoring allocations affect the
result.

Sensitivity also depended on the censoring mechanism. Administrative censoring
produced relatively little variation because censoring was concentrated near
the administrative time, whereas more dispersed censoring allowed the
allocation rules to differ across more intervals. These differences were most
apparent under coarser risk-table grids.

Table~\ref{tab:drop-density-diagnostics} summarizes the interval structure
associated with different risk-table resolutions. Coarser grids contained more
candidate drop times within each interval. When censoring was also present in
these intervals, this provided more possible within-interval censoring
allocations. The table reports the number of candidate drop times per interval
and the proportions of intervals in which alternative censoring allocations
can affect the reconstruction. The increase in such intervals with coarser
grids was clearer under oracle totals. Under practical totals, reconstruction
of \((D_k,C_k)\) resulted in fewer intervals satisfying \(J_k\ge 2\) and
\(C_k>0\), corresponding to intervals with multiple candidate drop times and a
positive reconstructed censoring total. Definitions of the drop-density and
allocation-relevance diagnostics are given in Supplementary Section~S5.3.
Additional method-specific reconstruction summaries under practical totals are
reported in Supplementary Section~S5.4.

\section{Numerical Illustration}
\label{sec:numerical-illustration}

We illustrate the reconstruction using the nivolumab plus ipilimumab arm from
the progression-free survival panel of Figure~2A in the 6.5-year CheckMate~067
report \citep{Wolchok2022CheckMate067}. This arm included \(n=314\) patients.
The Kaplan--Meier curve was extracted from the vector graphics in the
published PDF, and the number-at-risk table provided the interval-level
constraints. The published curve did not include censoring marks. Consider the
first risk-table interval \(I_0^{(3)}=[0,3)\), for which the reported numbers
at risk were
$$
Y_0^{(3)}=314,\qquad Y_1^{(3)}=218.
$$
Thus,
$$
D_0^{(3)}+C_0^{(3)}=96.
$$
Applying the Guyot reconstruction gives
$$
D_0^{\mathrm{practical},(3)}=61,\qquad
C_0^{\mathrm{practical},(3)}=35.
$$
The extracted Kaplan--Meier curve contains 21 candidate drop times within this
interval.

We then applied the working censoring profiles while keeping the event total,
censoring total, candidate drop times, and digitized drop fractions fixed. For
example, under the linear profile,
$$
q_{0,\mathrm{lin}}(t)=\frac{t}{3},
$$
the 35 censoring observations are allocated across the local segments
according to their lengths. The resulting cumulative censoring counts
\(M_{0j,\mathrm{lin}}\) determine the event-allocation weights
$$
w_{0j,\mathrm{lin}}
=
\left(314-M_{0j,\mathrm{lin}}\right)
\widehat h_{0j}^{(3)}.
$$
The fixed total of 61 events is then distributed across the 21 candidate drop
times according to these weights.

Different working profiles can change the cumulative censoring counts at the
candidate drop times and therefore the reconstructed event allocation. All
profiles nevertheless preserve the same interval totals and the reported
endpoint number at risk,
$$
314-61-35=218.
$$
The published curve fixes the candidate drop times, and the first-stage
reconstruction fixes the interval totals, while the within-interval censoring
allocation remains unspecified. Different working allocations can therefore
produce different reconstructed pseudo-IPD while satisfying the same
published constraints.
Figure~\ref{fig:checkmate067-profile-comparison} compares the working profiles
in the first interval. Panel~A shows that the refitted Kaplan--Meier curves
remain close to the digitized curve. Panel~B shows substantial differences in
cumulative censoring across profiles, although all profiles preserve the same
interval-level censoring total
\(C_0^{\mathrm{practical},(3)}=35\).

This example illustrates the many-to-one nature of the reconstruction. The
candidate Kaplan--Meier drop times and interval totals are fixed, but the
published information does not determine a unique within-interval censoring
allocation. Different working profiles can therefore produce different
cumulative censoring patterns and event allocations while preserving the same
reported endpoint constraint. Additional calculations for this example are
given in Supplementary Section~S6.

\section{Discussion and conclusion}
\label{sec:discussion-conclusion}

This study shows that pseudo-IPD reconstruction from a published
Kaplan--Meier curve and number-at-risk table can remain non-identifiable even
after the interval-level event and censoring totals have been determined.
The remaining uncertainty arises from the placement of censoring within each
risk-table interval. Our theoretical and numerical results show that different
within-interval censoring allocations can lead to different reconstructed risk
sets and event allocations while satisfying the same interval-level
constraints. Agreement with the published Kaplan--Meier curve is therefore an
important reconstruction check, but it does not establish that the underlying
individual-level follow-up pattern has been uniquely recovered. This issue is
particularly relevant for analyses that depend on the reconstructed risk sets
or censoring distribution, including IPCW-based and other censoring-sensitive
procedures.

The working profiles provide a simple way to examine the effect of this
unidentified within-interval censoring placement. They are not intended to
estimate the true censoring distribution. Instead, they generate alternative
within-interval allocations that satisfy the same interval-level censoring
total. If a downstream analysis gives similar results across these
reconstructions, the result is relatively insensitive to the working
allocation. Larger differences indicate greater dependence on the
within-interval censoring placement.

Optimization-based reconstruction methods such as
\citet{Titman2026QP} can be used to select a reconstruction satisfying the
published constraints. The present analysis addresses the variation that can
remain after interval-level totals have been fixed. Different within-interval
censoring allocations can then be carried forward through the same
reconstruction and downstream analysis. This approach can be used with
different first-stage reconstruction methods whenever the interval-level
totals and candidate event times are available.

Risk-table resolution also affects the extent of this variation. Wider
reporting intervals may contain more candidate drop times and, when censoring
is present, allow more possible within-interval censoring patterns.
Finer risk tables or identifiable censoring marks can reduce the range of
possible allocations. More detailed reporting of numbers at risk and
censoring information would therefore improve the usefulness of published
survival curves for secondary analysis.

In practice, researchers need not rely on a single reconstructed pseudo-IPD
dataset. Reconstructions generated under different within-interval censoring
allocations can be carried forward into the same downstream analysis.
Comparing the resulting estimates and conclusions provides a direct assessment
of how strongly the analysis depends on the reconstruction. More generally,
published Kaplan--Meier information may support multiple pseudo-IPD
reconstructions when within-interval censoring is not identified. Examining
results across these reconstructions provides a broader assessment of
reconstruction uncertainty than agreement with the published survival curve
alone.

\clearpage


\clearpage
\section*{Tables}

\begin{table}[htbp]
\centering
\caption{Simulation settings.}
\label{tab:simulation-settings}
\small
\begin{tabular}{ll}
\toprule
Item & Setting \\
\midrule
Monte Carlo replications & 500 per scenario \\
Sample size & 300 \\
Event-time distribution & Weibull, scale \(=12\) \\
Event shape values & \(0.8, 1.2, 2.0\) \\
Censoring mechanisms
& uniform, exponential, Weibull, administrative, mixture \\
Risk-table grid widths & \(3, 6, 12\) months \\
Total-count settings & oracle totals; practical totals \\
Working allocation profiles
& Guyot-style even, linear, front-loaded, back-loaded \\
Evaluation grid
& risk-table times plus Kaplan--Meier drop times \\
\bottomrule
\end{tabular}
\end{table}

\begin{table}[htbp]
\centering
\caption{Mean sensitivity ranges by risk-table grid width.}
\label{tab:simulation-grid-ranges}
\small
\setlength{\tabcolsep}{3pt}
\renewcommand{\arraystretch}{1.1}
\begin{tabular}{lcrrrr}
\toprule
\shortstack[l]{Total-count\\setting}
& \shortstack{Grid\\width}
& \shortstack{RMST\\error}
& \shortstack{Survival\\MAE}
& \shortstack{Risk-set\\MAE}
& \shortstack{Censoring\\MAE} \\
\midrule
Oracle totals    & 3  & 0.177 & 0.004 & 2.436  & 2.355 \\
Oracle totals    & 6  & 0.337 & 0.009 & 5.509  & 5.373 \\
Oracle totals    & 12 & 0.534 & 0.024 & 18.843 & 12.490 \\
\addlinespace
Practical totals & 3  & 0.177 & 0.004 & 2.434  & 2.352 \\
Practical totals & 6  & 0.319 & 0.007 & 5.415  & 5.353 \\
Practical totals & 12 & 0.333 & 0.004 & 11.239 & 10.870 \\
\bottomrule
\end{tabular}
\end{table}

\begin{table}[ht]
\centering
\caption{Drop-density and censoring-allocation diagnostics by risk-table grid width.}
\label{tab:drop-density-diagnostics}

\setlength{\tabcolsep}{9pt}
\renewcommand{\arraystretch}{1.15}

\begin{tabular}{cccccc}
\hline
Grid width & $\bar J$ & Mean $J_{\max}$ & $p_{\ge2}$
& \multicolumn{2}{c}{$p_{\mathrm{alloc}}$} \\
\cline{5-6}
& & & & Oracle & Practical \\
\hline
3  & 22.007 & 61.456  & 0.851 & 0.623 & 0.618 \\
6  & 43.502 & 102.775 & 0.907 & 0.686 & 0.661 \\
12 & 84.954 & 157.141 & 0.924 & 0.703 & 0.644 \\
\hline
\end{tabular}
\end{table}

\clearpage
\section*{Figures}

\begin{figure}[htbp]
\centering
\includegraphics[width=0.8\textwidth]{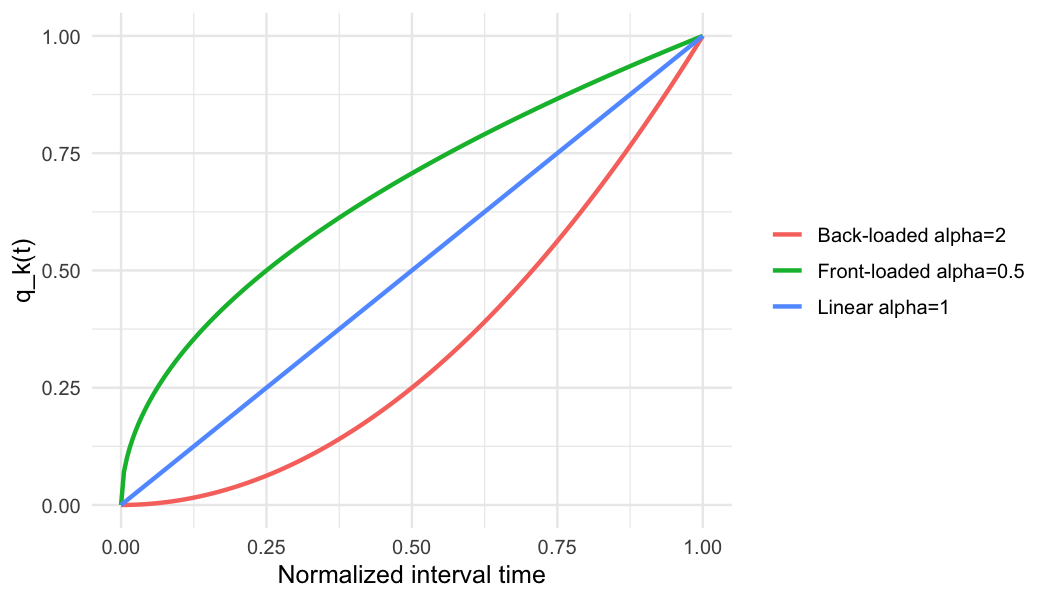}
\caption{Examples of within-interval working censoring profiles.
The profiles allocate the same interval-level censoring total \(C_k\), but place
it earlier, evenly, or later within the interval.}
\label{fig:working-profiles}
\end{figure}

\begin{figure}[htbp]
\centering
\includegraphics[width=\textwidth]{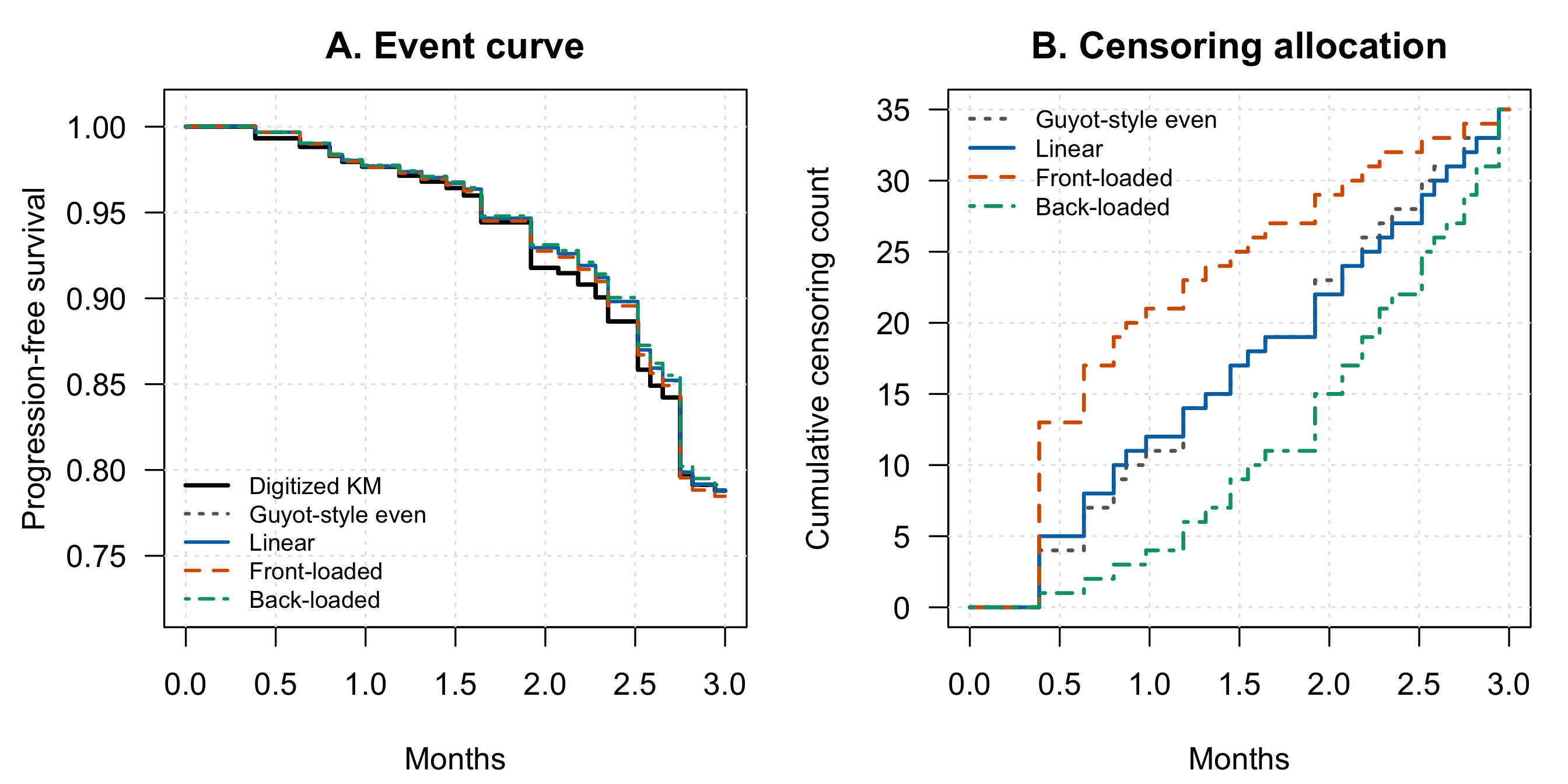}
\caption{Comparison of working allocation profiles in the first 0--3 month
risk-table interval of the CheckMate~067 nivolumab plus ipilimumab
progression-free survival curve. Panel~A compares the digitized Kaplan--Meier
curve with refitted pseudo-IPD curves under alternative within-interval
censoring profiles. Panel~B shows the corresponding cumulative censoring
allocations. The reconstructed event curves are similar across profiles,
whereas the implied censoring allocations differ visibly.}
\label{fig:checkmate067-profile-comparison}
\end{figure}

\clearpage

\setcounter{section}{0}
\setcounter{equation}{0}
\setcounter{table}{0}
\setcounter{figure}{0}
\setcounter{proposition}{0}
\setcounter{footnote}{0}
\renewcommand{\thesection}{S\arabic{section}}
\renewcommand{\theequation}{S\arabic{equation}}
\renewcommand{\thetable}{S\arabic{table}}
\renewcommand{\thefigure}{S\arabic{figure}}
\renewcommand{\theproposition}{S\arabic{proposition}}
\renewcommand{\theHsection}{S\arabic{section}}
\renewcommand{\theHequation}{S\arabic{equation}}
\renewcommand{\theHtable}{S\arabic{table}}
\renewcommand{\theHfigure}{S\arabic{figure}}
\renewcommand{\theHproposition}{S\arabic{proposition}}

\makeatletter
\let\maketitle\KMsaved@maketitle
\let\@maketitle\KMsaved@@maketitle
\let\@author\KMsaved@author
\gdef\@title{Supplementary Information for\\
Pseudo-IPD Reconstruction from Published Kaplan--Meier Curves under Unidentified Within-Interval Censoring}
\gdef\@date{}
\makeatother

\setstretch{1.12}
\maketitle
\doublespacing

\section{Overview}

This Supplement provides additional technical details, proofs, simulation
specifications, and numerical calculations supporting the main-text analyses.
Section~\ref{sec:supp-guyot} gives the first-stage Guyot-style reconstruction
used to obtain the interval-level totals. Section~\ref{sec:supp-integer-allocation}
gives the largest-remainder integer-allocation rule and finite-sample bounds.
Section~\ref{sec:supp-proofs} gives proofs of the theoretical results stated in
the main text. Section~\ref{sec:supp-simulation} provides the simulation
details, including the data-generating mechanisms, reconstruction metrics,
drop-density and censoring‑allocation diagnostics, and additional simulation
results. Section~\ref{sec:supp-checkmate067} gives additional calculations for
the CheckMate~067 example. Notation follows the main text unless a quantity is
introduced locally for an implementation detail.

\section{Guyot-style first-stage reconstruction}
\label{sec:supp-guyot}

This section gives the calculations underlying the first-stage reconstruction
summarized in the main text. For a risk-table interval
\(I_k=[t_k,t_{k+1})\), the digitized Kaplan--Meier curve and the reported
numbers at risk are used to obtain the interval-level event and censoring
totals \((D_k,C_k)\).

The number at risk that would remain at \(t_{k+1}\) in the absence of
within-interval censoring is estimated by
\begin{equation}
\widehat Y_{k+1}^{\mathrm{no\ censor}}
=
\operatorname{round}\left\{
Y_k
\frac{\widehat S_{\mathrm{dig}}(t_{k+1}^-)}
     {\widehat S_{\mathrm{dig}}(t_k)}
\right\}.
\label{eq:supp-guyot-no-censor-risk}
\end{equation}
The initial censoring estimate is
\begin{equation}
\widetilde C_k
=
\max\left\{
0,\,
\widehat Y_{k+1}^{\mathrm{no\ censor}}-Y_{k+1}
\right\}.
\label{eq:supp-guyot-initial-censor}
\end{equation}
The truncation at zero prevents negative censoring counts caused by
digitization or rounding discrepancies.
Given \(\widetilde C_k\), working censoring times are placed approximately
evenly within the interval:
\begin{equation}
\xi_{kc}
=
t_k+
\frac{c}{\widetilde C_k+1}(t_{k+1}-t_k),
\qquad
c=1,\ldots,\widetilde C_k.
\label{eq:supp-guyot-even-times}
\end{equation}
At candidate drop time \(u_{kj}\), the reconstructed number at risk immediately
before the drop is
\begin{equation}
R_{kj}
=
Y_k
-
\sum_{c=1}^{\widetilde C_k}I(\xi_{kc}<u_{kj})
-
\sum_{\ell=1}^{j-1}d_{k\ell},
\qquad
j=1,\ldots,J_k.
\label{eq:supp-guyot-local-risk}
\end{equation}
The event count at that drop is then
\begin{equation}
d_{kj}
=
\operatorname{round}\left\{
R_{kj}\widehat h_{kj}
\right\},
\qquad
j=1,\ldots,J_k.
\label{eq:supp-guyot-event-count}
\end{equation}

Because the reconstruction uses digitized survival values and integer event
counts, the initial censoring estimate need not reproduce the reported number
at risk exactly at \(t_{k+1}\). The candidate censoring total is therefore
adjusted, if necessary, and the sequential reconstruction is repeated until
\begin{equation}
Y_k-\widetilde C_k-\sum_{j=1}^{J_k}d_{kj}=Y_{k+1}.
\label{eq:supp-guyot-endpoint-match}
\end{equation}
After endpoint matching,
\[
C_k=\widetilde C_k,
\qquad
D_k=\sum_{j=1}^{J_k}d_{kj},
\]
so that \(D_k+C_k=Y_k-Y_{k+1}\).

The approximately even placement in
Equation~\eqref{eq:supp-guyot-even-times} is the working allocation used by
the first-stage Guyot-style reconstruction. The main-text method keeps the
resulting interval totals fixed and varies the within-interval censoring
allocation.

\section{Integer allocation and finite-sample bounds}
\label{sec:supp-integer-allocation}

The main text defines the real-valued censoring allocations
\(\Delta_{kj,m}^{*}\) and event allocations \(d_{kj,m}^{*}\), and states that
both are converted to integer counts using a deterministic largest-remainder
rule. This section gives the details of that rule. Here
\(\lfloor x\rfloor\) denotes the greatest integer less than or equal to \(x\).

\subsection{Integer allocation of censoring counts}

For \(j=1,\ldots,J_k+1\), define
\[
b_{kj,m}^{C}
=
\left\lfloor\Delta_{kj,m}^{*}\right\rfloor,
\qquad
r_{kj,m}^{C}
=
\Delta_{kj,m}^{*}-b_{kj,m}^{C},
\]
and
\[
R_{k,m}^{C}
=
C_k-\sum_{j=1}^{J_k+1}b_{kj,m}^{C}.
\]
Let \(\mathcal J_{k,m}^{C}\) denote the indices of the
\(R_{k,m}^{C}\) largest fractional remainders \(r_{kj,m}^{C}\), with ties
resolved by a fixed deterministic rule. Then
\begin{equation}
c_{kj,m}
=
b_{kj,m}^{C}
+
I(j\in\mathcal J_{k,m}^{C}),
\qquad
j=1,\ldots,J_k+1,
\label{eq:supp-censor-integer-allocation}
\end{equation}
and
\begin{equation}
\sum_{j=1}^{J_k+1}c_{kj,m}=C_k.
\label{eq:supp-censor-total}
\end{equation}

\subsection{Integer allocation of event counts}

For \(j=1,\ldots,J_k\), define
\[
b_{kj,m}^{D}
=
\left\lfloor d_{kj,m}^{*}\right\rfloor,
\qquad
r_{kj,m}^{D}
=
d_{kj,m}^{*}-b_{kj,m}^{D},
\]
and
\[
R_{k,m}^{D}
=
D_k-\sum_{j=1}^{J_k}b_{kj,m}^{D}.
\]
Let \(\mathcal J_{k,m}^{D}\) denote the indices of the
\(R_{k,m}^{D}\) largest fractional remainders \(r_{kj,m}^{D}\), with ties
resolved by the same deterministic rule. Then
\begin{equation}
d_{kj,m}
=
b_{kj,m}^{D}
+
I(j\in\mathcal J_{k,m}^{D}),
\qquad
j=1,\ldots,J_k,
\label{eq:supp-event-integer-allocation}
\end{equation}
and
\begin{equation}
\sum_{j=1}^{J_k}d_{kj,m}=D_k.
\label{eq:supp-event-total}
\end{equation}
Thus the reported endpoint constraint is retained:
\[
Y_k
-
\sum_{j=1}^{J_k}d_{kj,m}
-
\sum_{j=1}^{J_k+1}c_{kj,m}
=
Y_{k+1}.
\]

\subsection{Finite-sample bounds for integer-allocation error}

The numerical discrepancy introduced by integer allocation is distinct from
variation caused by the choice of within-interval censoring profile. Define the
cumulative real-valued censoring allocation by
\[
M_{kj,m}^{*}
=
\sum_{\ell=1}^{j}\Delta_{k\ell,m}^{*}
=
C_kq_{k,m}(u_{kj}),
\qquad
j=1,\ldots,J_k.
\]

\begin{proposition}
For \(j=1,\ldots,J_k+1\),
\begin{equation}
\left|c_{kj,m}-\Delta_{kj,m}^{*}\right|<1.
\label{eq:supp-local-rounding-bound}
\end{equation}
Consequently,
\begin{equation}
\left|M_{kj,m}-M_{kj,m}^{*}\right|<j,
\qquad
j=1,\ldots,J_k.
\label{eq:supp-cumulative-rounding-bound}
\end{equation}
If \(C_k>0\),
\begin{equation}
\max_{1\le j\le J_k}
\left|
\frac{M_{kj,m}}{C_k}
-
\frac{M_{kj,m}^{*}}{C_k}
\right|
<
\frac{J_k}{C_k}.
\label{eq:supp-normalized-rounding-bound}
\end{equation}
\end{proposition}

\begin{proof}
Under the largest-remainder rule, \(c_{kj,m}\) is either
\(\lfloor\Delta_{kj,m}^{*}\rfloor\) or
\(\lfloor\Delta_{kj,m}^{*}\rfloor+1\), which gives
Equation~\eqref{eq:supp-local-rounding-bound}. Also,
\[
M_{kj,m}-M_{kj,m}^{*}
=
\sum_{\ell=1}^{j}
\left(c_{k\ell,m}-\Delta_{k\ell,m}^{*}\right).
\]
Equation~\eqref{eq:supp-cumulative-rounding-bound} follows from the triangle
inequality. When \(C_k>0\), division by \(C_k\) and maximization over \(j\)
give Equation~\eqref{eq:supp-normalized-rounding-bound}. If \(C_k=0\), all
censoring allocations are zero.
\end{proof}
The same argument gives
\[
\left|d_{kj,m}-d_{kj,m}^{*}\right|<1,
\qquad
j=1,\ldots,J_k,
\]
while the total number of reconstructed events remains exactly \(D_k\).

\section{Proofs of theoretical results}
\label{sec:supp-proofs}

This section provides proofs of Proposition~1, Corollary~1,
Proposition~2, and Proposition~3 in the main text.

\subsection{Event-grid equivalence and within-segment invariance}

For Proposition~1, suppose
\[
\mathbf M_{k,m}=\mathbf M_{k,m'}.
\]
Then, at every candidate drop time,
\[
\widetilde Y_{kj,m}
=
Y_k-M_{kj,m}
=
Y_k-M_{kj,m'}
=
\widetilde Y_{kj,m'}.
\]
Because the digitized drop fractions \(\widehat h_{kj}\) are fixed,
\[
w_{kj,m}
=
\widetilde Y_{kj,m}\widehat h_{kj}
=
\widetilde Y_{kj,m'}\widehat h_{kj}
=
w_{kj,m'}.
\]
Hence \(W_{k,m}=W_{k,m'}\), and therefore
\[
d_{kj,m}^{*}
=
D_k\frac{w_{kj,m}}{W_{k,m}}
=
D_k\frac{w_{kj,m'}}{W_{k,m'}}
=
d_{kj,m'}^{*}.
\]
Applying the same deterministic integer-allocation rule gives the same integer
event counts.

For Corollary~1, changing the exact censoring times within a segment while
keeping its censoring count fixed does not change any cumulative censoring
count \(M_{kj,m}\) at the candidate drop times. The result therefore follows
from Proposition~1.

\subsection{Number of possible censoring patterns}

For Proposition~2, the segment-specific censoring counts satisfy
\[
c_{k1}+\cdots+c_{k,J_k+1}=C_k,
\qquad
c_{kj}\ge0.
\]
Thus, the possible censoring patterns are the nonnegative integer solutions to
this equation. By the stars-and-bars argument, their number is
\[
\binom{C_k+(J_k+1)-1}{(J_k+1)-1}
=
\binom{C_k+J_k}{J_k}.
\]

\subsection{Endpoint preservation}

For Proposition~3, if
\[
\sum_{j=1}^{J_k}d_{kj,m}=D_k,
\qquad
\sum_{j=1}^{J_k+1}c_{kj,m}=C_k,
\]
then
\[
Y_k
-
\sum_{j=1}^{J_k}d_{kj,m}
-
\sum_{j=1}^{J_k+1}c_{kj,m}
=
Y_k-D_k-C_k.
\]
Since
\[
D_k+C_k=Y_k-Y_{k+1},
\]
the right-hand side equals \(Y_{k+1}\), which proves the result.

\section{Simulation details}
\label{sec:supp-simulation}

This section gives the simulation details used to reproduce the main-text
analysis. The oracle-total and practical-total settings, the four allocation
rules, and the sensitivity measure \(R_{Q,a}^{(g)}\) follow the definitions
in the main text.

\subsection{Data-generating mechanisms}

Event times were generated from
\[
T_i
\sim
\operatorname{Weibull}
\left(
\text{shape}=\kappa_T,\,
\text{scale}=12
\right),
\qquad
\kappa_T\in\{0.8,1.2,2.0\}.
\]
Censoring times were generated from one of five mechanisms:
\[
\begin{aligned}
\text{Uniform:}\qquad
&C_i\sim\operatorname{Uniform}(0,24),\\
\text{Exponential:}\qquad
&C_i\sim\operatorname{Exponential}(\text{rate}=1/18),\\
\text{Weibull:}\qquad
&C_i\sim\operatorname{Weibull}(\text{shape}=1.4,\text{scale}=20),\\
\text{Administrative:}\qquad
&C_i=24,\\
\text{Mixture:}\qquad
&C_i=\min(C_i^{\mathrm{rand}},24),
\qquad
C_i^{\mathrm{rand}}\sim\operatorname{Uniform}(0,30).
\end{aligned}
\]
The observed data were \((X_i,\delta_i)\), as defined in the main text. Each
simulated dataset contained \(n=300\) subjects, and 500 Monte Carlo
replications were generated for each combination of event-time and censoring
distributions.

For grid width \(g\in\{3,6,12\}\), reporting times were
\begin{equation}
t_k=kg,
\qquad
k=0,1,\ldots,K,
\label{eq:supp-simulation-risk-grid}
\end{equation}
where
\begin{equation}
K
=
\left\lfloor\frac{\max_i X_i}{g}\right\rfloor+1.
\label{eq:supp-simulation-grid-end}
\end{equation}
For each grid width, the corresponding numbers at risk, candidate drop times,
and digitized drop fractions were obtained from the fitted Kaplan--Meier curve
and risk table. Candidate drop times and drop fractions were obtained directly
from the fitted curve, so no additional curve-digitization error was introduced.

\subsection{Reconstruction metrics}

For the Kaplan--Meier curve, number-at-risk process, and cumulative censoring
process, \(Q_{a,m}^{(g)}\) was calculated as the mean absolute difference
between the reconstructed and original observed quantities over the combined
set of risk-table times and candidate Kaplan--Meier drop times.
For RMST, \(\tau\) was the 90th percentile of observed follow-up times in each
simulated dataset, as in the main text. The same \(\tau\) was used for the
original data and all reconstructions of that dataset, and
\(Q_{a,m}^{(g)}\) was the absolute difference between the reconstructed and
original observed RMST.
The across-profile range \(R_{Q,a}^{(g)}\) was calculated as defined in the
main text. The ranges were interpreted separately for each reconstruction
quantity because the quantities are on different scales.

\subsection{Drop-density and allocation-relevance diagnostics}

For each simulated dataset and grid width, \(K\) denotes the number of
risk-table intervals. The mean and maximum numbers of candidate drop times per
interval were
\begin{equation}
\bar J
=
\frac{1}{K}\sum_{k=0}^{K-1}J_k,
\qquad
J_{\max}
=
\max_{0\le k\le K-1}J_k.
\label{eq:supp-drop-density}
\end{equation}
The proportion of intervals containing at least two candidate drop times was
\begin{equation}
p_{\ge2}
=
\frac{1}{K}
\sum_{k=0}^{K-1}I(J_k\ge2),
\label{eq:supp-multiple-drop-proportion}
\end{equation}
and the proportion of allocation-relevant intervals was
\begin{equation}
p_{\mathrm{alloc}}
=
\frac{1}{K}
\sum_{k=0}^{K-1}I(J_k\ge2,\ C_k>0).
\label{eq:supp-allocation-relevant-proportion}
\end{equation}
For practical totals, \(C_k\) is the reconstructed interval censoring total.
For oracle totals, it is the corresponding observed interval censoring total.
The values reported in the main-text table are Monte Carlo averages of these
dataset-level quantities.

\subsection{Additional simulation results}

Table~\ref{tab:supp-method-specific-practical} gives the average reconstruction
errors for the four allocation rules under practical totals. The relative
performance of the allocation rules varied across the four reconstruction
quantities. 

\begin{table}[htbp]
\centering
\caption{Method-specific reconstruction summaries under practical totals.}
\label{tab:supp-method-specific-practical}
\small
\setlength{\tabcolsep}{4pt}
\begin{tabular}{lrrrr}
\toprule
Allocation profile
& \shortstack{RMST\\error}
& \shortstack{Survival\\MAE}
& \shortstack{Risk-set\\MAE}
& \shortstack{Censoring\\MAE} \\
\midrule
Guyot-style even & 0.106 & 0.011 & 7.352 & 9.829 \\
Linear profile & 0.103 & 0.011 & 7.710 & 10.139 \\
Front-loaded \(\alpha=0.5\) & 0.206 & 0.014 & 6.606 & 9.376 \\
Back-loaded \(\alpha=2\) & 0.158 & 0.013 & 11.986 & 14.289 \\
\bottomrule
\end{tabular}
\end{table}

\section{CheckMate 067 additional calculations}
\label{sec:supp-checkmate067}

We give additional calculations for the nivolumab plus ipilimumab arm from the
progression-free survival panel of Figure~2A in the 6.5-year CheckMate~067
report \citep{Wolchok2022CheckMate067S}. The Kaplan--Meier curve was extracted
from the vector path in the published PDF. The published figure did not show
censoring marks, so the reconstruction used the survival path and the
number-at-risk table. The extracted curve was compared with an available
Kaplan--Meier curve derived from individual patient data on a 0.25-month grid
through 73.5 months. The mean and maximum absolute curve differences were
0.900 and 4.075 percentage points, respectively. These differences quantify
curve-extraction error and are separate from the within-interval censoring
allocation analysis.

For the first risk-table interval, \(I_0^{(3)}=[0,3)\), the reported numbers
at risk were \(Y_0^{(3)}=314\) and \(Y_1^{(3)}=218\). The practical
reconstruction gives \(D_0^{\mathrm{practical},(3)}=61\) and
\(C_0^{\mathrm{practical},(3)}=35\). The extracted curve contains
\(J_0^{(3)}=21\) candidate Kaplan--Meier drop times. These fixed totals and
drop times are used below to illustrate the linear working profile. For
readability, the practical-total and linear-profile labels are suppressed in
the remaining notation. Thus, \(D_0=61\), \(C_0=35\), and
\(M_{0j}\), \(\widetilde Y_{0j}\), \(w_{0j}\), and \(d_{0j}^{*}\)
denote the corresponding quantities for this reconstruction.

For the linear profile, \(q_{0,\mathrm{lin}}(t)=t/3\). With
\(u_{0,0}^{(3)}=0\) and \(u_{0,22}^{(3)}=3\), the 35 censoring observations are
allocated over 22 segments according to
\begin{equation}
\Delta_{0j}^{*}
=
35\left\{
q_{0,\mathrm{lin}}(u_{0j}^{(3)})
-
q_{0,\mathrm{lin}}(u_{0,j-1}^{(3)})
\right\},
\qquad
j=1,\ldots,22.
\label{eq:supp-checkmate-segment-allocation}
\end{equation}
The largest-remainder rule gives integer censoring counts \(c_{0j}\) satisfying
\(\sum_{j=1}^{22}c_{0j}=35\). For each candidate drop,
\[
M_{0j}
=
\sum_{\ell=1}^{j}c_{0\ell},
\qquad
\widetilde Y_{0j}
=
314-M_{0j},
\qquad
j=1,\ldots,21.
\]
The corresponding real-valued cumulative censoring allocation is
\(M_{0j}^{*}=35q_{0,\mathrm{lin}}(u_{0j}^{(3)})\). Censoring in the final
segment \([u_{0,21}^{(3)},3)\) contributes to the interval total of 35 but not
to the cumulative censoring counts at the candidate drop times.
The event-allocation weights are
\[
w_{0j}
=
\widetilde Y_{0j}\widehat h_{0j}^{(3)},
\qquad
W_0
=
\sum_{j=1}^{21}w_{0j}
=
62.4059.
\]
The fixed event total is allocated according to
\[
d_{0j}^{*}
=
61\frac{w_{0j}}{62.4059},
\qquad
j=1,\ldots,21.
\]
The largest-remainder rule gives integer event counts \(d_{0j}\) satisfying
\(\sum_{j=1}^{21}d_{0j}=61\). Hence the reported endpoint is reproduced:
\(314-61-35=218\).

\begin{figure}[htbp]
\centering
\includegraphics[width=0.82\textwidth]
{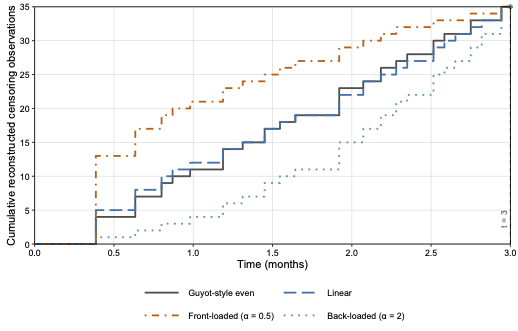}
\caption{Cumulative reconstructed censoring allocation during the first
risk-table interval, 0--3 months, in the CheckMate~067 example. All working
profiles have the same interval-level censoring total \(C_0=35\) but differ in
the timing of censoring within the interval.}
\label{fig:supp-checkmate-censoring-allocation}
\end{figure}

Table~\ref{tab:supp-checkmate067-linear-allocation} gives the first 10
candidate drop times and the corresponding allocation quantities. Figure~\ref{fig:supp-checkmate-censoring-allocation} compares the cumulative
censoring allocations produced by the four working profiles in the first
risk-table interval. All profiles reach the same interval total
\(C_0=35\) at 3 months, but they differ in how censoring accumulates within
the interval. The front-loaded profile places more censoring before the early
candidate drop times, whereas the back-loaded profile places more censoring
later in the interval. The linear and Guyot-style even allocations lie between
these two patterns. These differences change the cumulative censoring counts
available at the candidate drop times and can therefore change the event
allocation, even though the interval-level censoring total is fixed.

\clearpage

\begin{table}[htbp]
\centering
\scriptsize
\caption{Linear-profile censoring allocation and event weighting for the first
10 candidate Kaplan--Meier drops in the 0--3 month interval of the
CheckMate~067 nivolumab plus ipilimumab progression-free survival curve.}
\label{tab:supp-checkmate067-linear-allocation}
\resizebox{\textwidth}{!}{%
\begin{tabular}{rrrrrrrrr}
\toprule
\(j\) &
\(u_{0j}^{(3)}\) &
\(\widehat h_{0j}^{(3)}\) &
\(M_{0j}^{*}\) &
\(M_{0j}\) &
\(c_{0j}\) &
\(\widetilde Y_{0j}\) &
\(w_{0j}\) &
\(d_{0j}^{*}\) \\
\midrule
1  & 0.3860 & 0.004381 & 4.5034  & 5  & 5 & 309 & 1.3537 & 1.3234 \\
2  & 0.6346 & 0.005030 & 7.4039  & 8  & 3 & 306 & 1.5393 & 1.5049 \\
3  & 0.8004 & 0.005367 & 9.3376  & 10 & 2 & 304 & 1.6314 & 1.5949 \\
4  & 0.8697 & 0.003497 & 10.1467 & 11 & 1 & 303 & 1.0597 & 1.0359 \\
5  & 0.9805 & 0.002859 & 11.4392 & 12 & 1 & 302 & 0.8633 & 0.8440 \\
6  & 1.1877 & 0.002874 & 13.8563 & 14 & 2 & 300 & 0.8623 & 0.8430 \\
7  & 1.3120 & 0.003538 & 15.3066 & 15 & 1 & 299 & 1.0579 & 1.0342 \\
8  & 1.4503 & 0.003868 & 16.9196 & 17 & 2 & 297 & 1.1489 & 1.1232 \\
9  & 1.5471 & 0.004522 & 18.0493 & 18 & 1 & 296 & 1.3384 & 1.3084 \\
10 & 1.6439 & 0.016256 & 19.1790 & 19 & 1 & 295 & 4.7955 & 4.6881 \\
\bottomrule
\end{tabular}%
}
\end{table}

\clearpage

\end{document}